\documentclass[12pt]{article}
\usepackage{amsfonts}

\makeatletter
\@addtoreset{equation}{section} 
\makeatother

\newcommand{\eqn}[1]{(\ref{#1})}
\def\appendix#1
{
\addtocounter{section}{1}\setcounter{equation}{0}
 \renewcommand{\thesection}{\Alph{section}}
 \section*{Appendix~\thesection\protect\indent \parbox[t]{11.715cm}{#1}}
 \addcontentsline{toc}{section}{Appendix \thesection\ \ \ #1}
}

\def\<#1,#2>{\left\langle#1,#2\right\rangle} 

\def\nn{\nonumber}

\def\be{\begin{equation}}
\def\ee{\end{equation}}
\def\beqa{\begin{eqnarray}}
\def\eeqa{\end{eqnarray}}
\def\bd{\begin{displaymath}}
\def\ed{\end{displaymath}}

\def\del{\partial}

\newcommand{\bea}{\begin{array}}
\newcommand{\ea}{\end{array}}

\newcommand{\rh}{\rho}

\newcommand{\bean}{\begin{eqnarray*}}
\newcommand{\eean}{\end{eqnarray*}}
\newcommand{\he}{\hat{E}}
\newcommand{\hp}{\hat{p}}
\newcommand{\dd}[2]{ \delta_{ y_{ #1 },x_{ #2 } } }
\newcommand{\ddd}[2]{ \delta_{ y'_{ #1 }, x'_{ #2 } } }

\newcommand{\ap}{\alpha}
\newcommand{\bt}{\beta}
\newcommand{\hi}{\hat{I}}
\newcommand{\hpi}{\hat{\Pi}}
\newcommand{\pud}{\psi(x_1,x_2)}
\newcommand{\pdu}{\psi(x_2,x_1)}

\begin{document}
\begin{titlepage}
\begin{flushright}
\baselineskip=12pt
DSF-20/01\\
quant-ph/0202062\\
\hfill{ }\\
\end{flushright}

\begin{center}

\baselineskip=24pt

{\Large
\bf Permutation symmetry for the tomographic probability
distribution of a system of identical particles}

\baselineskip=14pt

\vspace{1cm}

{\bf V. I. Man'ko} $^{a,b}$, {\bf L. Rosa} $^{b}$,
 and {\bf P. Vitale} $^{c}$\\[6mm]
$^a$ {\it
P. N. Lebedev Physical Institute ,
Leninsky Pr. 53, Moscow, Russia
}\\[6mm]
$^b$ {\it Dipartimento di Scienze Fisiche, Universit\`{a} di Napoli {\sl
Federico II}\\
and {\it INFN, Sezione di Napoli, Monte S.~Angelo}\\
Via Cintia, 80126 Napoli, Italy}\\
{\tt manko@na.infn.it, luigi.rosa@na.infn.it}\\[6mm]
$^c$ {\it Dipartimento di Fisica, Universit\`{a} di Salerno and INFN
\\ Gruppo Collegato di Salerno, Via S. Allende
\\84081 Baronissi (SA), Italy}
\\ {\tt patrizia.vitale@sa.infn.it}\\

\end{center}

\vskip 2 cm

\begin{abstract}
The symmetry properties under permutation of tomograms representing the
states of a system of identical particles are studied. Starting from the
action of the permutation group on the density matrix we define its action
on the tomographic probability distribution. Explicit calculations are
performed  in the case of the two-dimensional harmonic oscillator.

{PACS: 03.65.Wj, 05.30.-d, 02.20.-a}
\end{abstract}
\end{titlepage}

\section{Introduction}

In the conventional approach to  quantum mechanics, that is the wave
function description \cite{Sc26}, the peculiarities  of the quantum
behaviour of a system of identical particles are encoded in the symmetry
properties of the wave functions. The description in terms of density
operators \cite{vN32} also implies, for the density matrices associated to
the states of identical particles, definite symmetry properties.
Nevertheless both the descriptions are very different from the  approach of
classical statistical mechanics in terms of probability distributions
defined on the phase space of the system. The wish to render the quantum
picture closer to the classical one gave rise to hidden variables theories
\cite{Bo52} and to the introduction of the Wigner quasi-distribution
function on phase space \cite{Wi32} obeying to the Moyal equation
\cite{Mo49}. The same aim of describing quantum states in terms of a
classical distribution of probability  determined then the introduction of
another class of quasi-distributions for the quantum states
\cite{Hu40,Su63, Gl63}.  In 1969 Cahill and Glauber \cite{CG69} considered
a set of s-ordered quasi-distribution functions that further generalized
the quasi-probability distribution functions previously introduced by
Wigner, Husimi, Glauber and Sudarshan.   They also showed that for each
quantum state one can find some probability distribution which is
determined by the density operator of the given state. But it was not clear
whether or not such probability distribution determined uniquely the
density operator, that is, whether or not the map was invertible. Only
recently it was realized that quantum mechanics could be described
completely in terms of this kind of probability distributions suitably
defined for a random variable \cite{MMT96}. In \cite{MMT96} a consistent
scheme has been proposed, the so-called {\em probability representation},
that results completely equivalent, in the sense of invertibility, to the
ordinary formulation of quantum mechanics: the quantum states are described
by a tomographic distribution of probability, or tomogram, also known as
marginal distribution function (MDF), and the evolution of the system is
described by an integro-differential equation for the MDF of  generalized
Fokker-Planck type. The probability representation of quantum mechanics 
uses as a mathematical tool the symplectic tomography map \cite{JA95} 
of density operators onto quadrature probability distributions.
 For a  general approach to tomograms and
quasi-distributions of quantum states see also \cite{BM99, MMV01} and, with
the inclusion of spin, \cite{AB01,DM97,MM97,Ag98}.

In the case of identical particles it is well known that they obey either
Bose or Fermi statistics. As a result the wave function describing the
system must be symmetrized or antisymmetrized respectively. Because the
states of a physical system belong to a vector space and are, consequently,
linearly superposable, it results quite easy to implement the two types of
statistics in the ordinary quantum mechanics considering, for example, the
wave functions of the system as a basis for a representation of the
permutation group. On the other hand, in the probability representation of
quantum states one needs to know what is the behaviour of the tomographic
probability distributions for systems of identical particles. The aim of
our article is the formulation of the symmetry properties of the tomograms
for systems of  identical particles.

In the following we use the relation between the MDF and the density matrix
to construct the action of the permutation group on the marginal
distribution functions. Because the set  of the MDFs is not a vector space
(see for example \cite{MMSZ00} where a superposition principle was
formulated for both the sets of density matrices and tomograms) we  obtain
a realization, and not a representation, of the permutation group on the
MDFs. In this way we can introduce the {\em completely symmetrized} and
{\em antisymmetrized} marginal distribution function.

The paper is organized as follows. In section 2 we review the realization
of the permutation group on the set of density matrices and in section 3 we
derive the corresponding realization on the set of MDFs.  For the sake of
clarity we work explicitly on the case of two identical particles and then
we extend the results to $n$ particles in section 4. In section 5 we
describe a simple application and finally the concluding remarks.

\section{The permutation group of the density matrix}
Identical particles obey either Bose or Fermi statistics. This property
implies two specific behaviours of the wave function of two identical
particles: it must be symmetric in the case of Bose-particles,
antisymmetric for Fermi-particles. These properties are more precisely
described by means of the permutation group representation theory. To have
a model let us concentrate on the system of two one dimensional particles
whose positions are $x_1$ and $x_2$ respectively. The corresponding wave
function,  $\psi(x_1,x_2)$, can be decomposed into the sum of a symmetric
and an antisymmetric function:
\beqa
& & \psi(x_1,x_2) =\frac{1}{2}(\psi(x_1,x_2)+\psi(x_2,x_1))+
\frac{1}{2}(\psi(x_1,x_2)-\psi(x_2,x_1))=  \nn \\
& & \psi_+(x_1,x_2)+\psi_-(x_1,x_2)
\label{eq:s1}
\eeqa
This decomposition may be related to the irreducible representation of the
permutation group $G=(\he,\hp_{12})$, where $\he$ is the identity of the
group and $\hp_{12}$ represents the operation of permutation of the
coordinates $x_1$ and $x_2$ i.e.
\beqa
\he\pud &=& \pud; \nn \\   
\hp_{12}\pud &=& \pdu. \label{eq:s4}
\eeqa
The table of characters of the representations of the permutation group
of two elements has the form
\beqa
\he & & \hp_{12} \nn \\
  1 & &  1 \\
  1 & &  -1 \nn
\eeqa
It means that the decomposition (\ref{eq:s1}) is connected with the
irreducible representations (\ref{eq:s4}) through the following formulas:
\beqa
\he\psi_\pm(x_1,x_2) &=& +1\psi_+(x_1,x_2)  \nn \\
\hp_{12}\psi_\pm(x_1,x_2) &=&\pm1\psi_+(x_1,x_2). \label{eq:s5}
\eeqa
Thus the two functions $\psi_+$ and $\psi_-$ realize a basis of the
one-dimensional representation of the permutation group. We say that
identical particles with symmetric wave function are described by means of
the symmetric representation of $G$, while identical particles with
antisymmetric wave function by means of the antisymmetric one.

The density matrix $\rh(x_1,x_2,x'_1,x'_2)$ of a pure state of two
particles with wave function $\pud$ has the form
\be
\rh(x_1,x_2,x'_1,x'_2)=\pud\psi^*(x_1',x_2') \label{eq:s7}
\ee
thus we can extend the action of the permutation group to the set of
density matrices because the coordinates $x_1,x_2$ and $x'_1,x'_2$ can be
permuted independently. This means that the group $\widetilde{G}=G\otimes
G$ which is the direct product of the permutation groups of two elements is
the one related to the symmetry properties of the density matrix.
$\widetilde{G}$ contains four elements:
\be
\hi=\he\otimes\he,~~\hat{P}_{12}=\hp_{12}\otimes\he,~~
\hat{P'}_{12}=\he\otimes\hp_{12},~~\hat{\Pi}_{12}=\hp_{12}
\otimes\hp_{12}. \label{eq:s8}
\ee
The action of the four elements on the density matrix (\ref{eq:s7}) is
defined as follows:
\beqa
\hi\rh(x_1,x_2,x'_1,x'_2) &=& \rh(x_1,x_2,x'_1,x'_2)  \label{eq:s9} \\
\hat{P}_{12}\rh(x_1,x_2,x'_1,x'_2) &=&\rh(x_2,x_1,x'_1,x'_2)\label{eq:s10} \\
\hat{P'}_{12}\rh(x_1,x_2,x'_1,x'_2)&=&\rh(x_1,x_2,x'_2,x'_1)\label{eq:s11} \\
\hat{\Pi}_{12}\rh(x_1,x_2,x'_1,x'_2)&=&\rh(x_2,x_1,x'_2,x'_1).\label{eq:s12}
\eeqa
Using the multiplication table of the group $\widetilde{G}$:
\begin{center}
\begin{tabular}{c|cccc}
 & $\hi$ & $\hat{P}_{12}$ & $\hat{P'}_{12}$ & $\hat{\Pi}_{12}$ \\ \hline
$\hi$ & $\hi$ & $\hat{P}_{12}$  & $\hat{P'}_{12}$  & $\hat{\Pi}_{12}$ \\
$\hat{P}_{12} $  & $\hat{P}_{12}$ & $\hi$ & $\hat{\Pi}_{12}$ &
$\hat{P'}_{12}$ \\
$\hat{P'}_{12} $ & $\hat{P'}_{12}$ & $\hat{\Pi}_{12}$ &
$\hi$ & $\hat{P}_{12}$ \\
$\hat{\Pi}_{12} $ & $\hat{\Pi}_{12}$ & $\hat{P'}_{12}$ &
$\hat{P}_{12}$ & $\hi$ \\
\end{tabular}
\end{center}
we may construct a realization through the action of $\widetilde{G}$ on the
following distributions:
\beqa
\rh_+ &=& \frac{1}{4}\left(
\hi\rh+\hpi\rh+\hat{P}_{12}\rh+\hat{P'}_{12}\rh  \right) \\
\rh_- &=& \frac{1}{4}\left(
\hi\rh+\hpi\rh-\hat{P}_{12}\rh-\hat{P'}_{12}\rh  \right) \\
\rh_1 &=& \frac{1}{4}\left(
\hi\rh-\hpi\rh+\hat{P}_{12}\rh-\hat{P'}_{12}\rh  \right) \\
\rh_2 &=& \frac{1}{4}\left(
\hi\rh-\hpi\rh-\hat{P}_{12}\rh+\hat{P'}_{12}\rh  \right).
\eeqa
It is easy to see that such an action is given by:
\begin{center}
\begin{tabular}{ccccc}
 & $\hi$ & $\hat{P}_{12}$ & $\hat{P'}_{12}$ & $\hat{\Pi}_{12}$ \\
$\rh_+$ & 1 & 1  & 1  & 1 \\ $\rh_-$ & 1 & -1 & -1 & 1 \\ $\rh_1$ & 1 & 1 &
-1 & -1 \\ $\rh_2$ & 1 & -1 & 1 & -1 .\\
\end{tabular}
\end{center}
Incidentally we note that $\rh_+$ and $\rh_-$ are also obtained as:
\beqa
& &\rh_+(x_1,x_2,x'_1,x'_2)=\psi_+(x_1,x_2)\psi_+^*(x_1',x_2')= 
\label{eq:s13}\\
& & \frac{1}{4}\left[(\psi(x_1,x_2)\psi^*(x_1',x_2')+
\psi(x_2,x_1)\psi^*(x_1',x_2')+
\psi(x_1,x_2)\psi^*(x_2',x_1')+
\psi(x_2,x_1)\psi^*(x_2',x_1') \right],  \nn
\eeqa
and
\beqa
& & \rh_-(x_1,x_2,x'_1,x'_2)=\psi_-(x_1,x_2)\psi_-^*(x_1',x_2')=  
\label{eq:s14}\\
& & \frac{1}{4}\left[ \psi(x_1,x_2)\psi^*(x_1',x_2')-
\psi(x_2,x_1)\psi^*(x_1',x_2')-
\psi(x_1,x_2)\psi^*(x_2',x_1')+
\psi(x_2,x_1)\psi^*(x_2',x_1') \right] \nn  
\eeqa
that is $\rho_+$ and $\rho_-$ are the usual density matrices respectively
associated to the usual symmetric and antisymmetric wave functions. (Here
and in the whole paper everything is  obtained for pure states but because
of the linearity of the mixtures of pure states we can easily generalise to
the density matrices of arbitrary mixed states of two particles.)

Thus, what we have done so far is to extend the action of the permutation
group (which we know for the space of the wave functions) to the space of
density matrices (Eqs. \eqn{eq:s9}-\eqn{eq:s12}) by means of Eq.
\eqn{eq:s7} which links in a simple way the density matrices to the wave
functions. The natural generalization of this procedure, that is the
definition of the action of the permutation group on the space of tomograms
through the expression of tomograms in terms of density matrices, is not so
simple since such a relation is an integral transform. This implies that
the density matrices and the MDFs depend on different variables and, as we
will see in the next section, it is not clear what is the action of the
permutation group on the new variables. We follow then a different approach
which we illustrate preliminarly for the space of the density matrices.
What we are looking for is an expression of the symmetrized and
antisymmetrized density matrices in integral form and, more precisely, we
search for a realization of the permutation group in terms of integral
kernels. Starting from Eqs.  \eqn{eq:s9}, \eqn{eq:s12} it is easy to
realize that
\beqa
\rho_\pm &=&\frac{1}{4}\left\{ \rh(x_1,x_2,x'_1,x'_2)  +
\rh(x_1,x_2,x'_2,x'_1)  \pm\left[\rh(x_2,x_1,x'_1,x'_2) +
\rh(x_1,x_2,x'_2,x'_1)  \right] \right\} \nn \\
&=&\frac{1}{4}\int dy_1dy_2dy'_1dy'_2
\left\{ \dd{1}{1}\dd{2}{2}\ddd{1}{1}\ddd{2}{2}+
\dd{1}{2}\dd{2}{1}\ddd{1}{2}\ddd{2}{1}+\right. \nn \\
&\pm  &\left.\left[
\dd{1}{2}\dd{2}{1}\ddd{1}{1}\ddd{2}{2}+
\dd{1}{1}\dd{2}{2}\ddd{1}{2}\ddd{2}{1} \right] \right\}
\rh(y_1,y_2,y'_1,y'_2) ,
\eeqa
with $\dd{i}{j}=\delta(y_i-x_j)$, that is
\be
\rh_\pm(x_1,x_2;x'_1,x'_2)=\int K_\pm(x_1,x_2,x'_1,x'_2;y_1,y_2,y'_1,y'_2)
\rh(y_1,y_2;y'_1,y'_2)dy_1dy_2dy'_1dy'_2, \label{eq:kern1}
\ee
with the kernel
\beqa
 K_\pm(x_1,x_2,x'_1,x'_2;y_1,y_2,y'_1,y'_2)&=&\frac{1}{4}\left\{
\dd{1}{1}\dd{2}{2}\ddd{1}{1}\ddd{2}{2}+
\dd{1}{2}\dd{2}{1}\ddd{1}{2}\ddd{2}{1}\pm \right.\nn \\
&&
\left.\left[\dd{1}{2}\dd{2}{1}\ddd{1}{1}\ddd{2}{2}+
\dd{1}{1}\dd{2}{2}\ddd{1}{2}\ddd{2}{1} \right]\right\}.
\eeqa
Defining the permutation operator as
\be
P_{j_1,j_2\ldots,j_n}f(x_1,x_2,\ldots,x_n)=
f(x_{j_1},x_{j_2},\ldots,x_{j_n})
\ee
we can write the kernel as:
\be
K_\pm(x,x'|y,y')_2=\frac{1}{(2!)^2}
(I\pm P_{12})\otimes
(I\pm P'_{12}) \dd{1}{1}\dd{2}{2}\ddd{1}{1}\ddd{2}{2}
\ee
where $(x,x'|y,y')_2\equiv (x_1,x_2,x'_1,x'_2;y_1,y_2,y'_1,y'_2)$ and we
assume that $P,~P'$ act on $x$, $x'$ respectively. Analogously we may
define the kernels $K_1,K_2$ associated to the density matrices
$\rho_1,\rho_2$.
 In this way we obtain  a realization of the permutation group
in terms of integral kernels. This approach is absolutely equivalent to the
one previously described but it is the one we need to implement the
symmetry properties in the MDF framework.

\section{The permutation group of the MDF}

We are now in the position to extend the previous procedure to the
tomographic probability distribution, but we first need a brief review
 of the theory.

 The MDF of a random variable $X$ is defined in \cite{CG69} as the Fourier
transform of the quantum characteristic function $\chi(k)=<e^{ik\hat{X}}>$:
\be
w(X,t)=\frac{1}{2\pi}\int{dk}e^{-ik{X}}<e^{ik\hat{X}}> \label{mdf}
\ee
where $\hat{X}$ is the operator associated to $X$, and, for each observable
$\hat{O},~~<\hat{O}>=Tr(\hat{\rho}\hat{O})$ with $\hat{\rho}$ the
time-dependent density operator. The MDF so defined is positive and 
normalized to unity, provided $\hat X$ is an observable \cite{CG69}.

In Ref. \cite{MMT95} it is shown that by taking 
\be
X=\mu q+\nu p, \label{x} 
\ee
with $q$
and $p$ two conjugate variables and $\mu,\nu$ real parameters labelling
different reference frames in phase space, $w(X,\mu,\nu,t)$ is normalized 
with respect to the $X$ variable and there exists an invertible
relation among the MDF and the density matrix. The variable $X$ represents
the position coordinate taking values in an ensemble of reference frames.
Equation \eqn{mdf} may be rewritten in the more convenient form 
\cite{MRV98b}
\be
w(X,\mu,\nu,t)={1\over 2\pi |\nu|\hbar} \int \rho(Z',Z'',t) \exp 
\left[-i{Z'-Z''\over \nu\hbar}\left(X-\mu {Z'+Z''\over 2}
\right)\right] dZ'\, dZ'' \label{wro}
\ee
and its inverse is represented by 
\be
\rho (X,X',t)={1\over 2\pi}\int w(Y,\mu, X-X') \exp \left[ 
{i\over \hbar} \left(Y-\mu{X+X'\over 2}\right)\right] d\mu \, dY \label{row}
\ee
It is important to note that, for \eqn{wro} to be invertible, it is 
necessary that $X$ be a coordinate variable taking values in an 
ensemble of phase 
spaces; in other words, the specific choice $\mu=1, \nu=0$ or any 
other fixing of the parameters $\mu$ and $\nu$ 
would not allow to reconstruct the density matrix.  
Hence, the MDF contains the same amount of information on a quantum 
state as the density matrix, only if Eq. \eqn{x} is assumed. 

For Hamiltonians of the form 
\be
H={p^2\over 2m} + V(q)
\ee  
an evolution equation governing the time dependence of the MDF is
available (see ref. \cite{MRV98b} for a simple derivation)
\beqa
\del_t w(X,\mu,\nu,t) &=& \left\{{\mu\over m} \del_\nu +{i\over \hbar}
[V(-(\del_X)^{-1}\del_\mu -{i\nu\hbar\over 2} \del_X) \right. \nn\\
&-& \left. V(-(\del_X)^{-1}\del_\mu +{i\nu\hbar\over 2} \del_X)]
\right\}w(X,\mu,\nu,t)  \label{evo}
\eeqa
where  the inverse derivative is defined as
\be
(\del_X)^{-1}\int f(Z) e^{g(Z) X} dZ = \int {f(Z)\over g(Z)}e^{g(Z) X} dZ.
\ee
The evolution equation, of generalized Fokker-Plank type, 
plays the r\^ole of the Schr\"odinger equation in 
the alternative scheme we are outlining. 
Its classical limit is easily seen to be 
\be
{\dot w}(X,\mu,\nu,t)= \left\{\frac{\mu}{m} {\del\over \del \nu}+\nu 
V'\left(-\left({\del \over \del X}\right)^{-1} {\del \over \del 
\mu}\right){\del \over \del X} \right\} w(X,\mu,\nu,t)~,
\label{evocl}
 \ee
where $V'$ is the derivative of the potential with respect to the 
argument. Equation \eqn{evocl} 
may be checked to be equivalent to Boltzmann equation 
for a classical distribution of probability $f(q,p,t)$ ,
\be
{\del f\over \del t} + {p\over m} {\del f \over \del q} -  {\del V \over 
\del q} {\del f \over \del p} =0,
\ee
after performing the change of variables
\be
w(X,\mu,\nu,t) = {1\over 2\pi} \int f(q,p,t) e^{ik(X-\mu q -\nu p) } 
dk~ dq~ dp~;
\ee
Hence, the classical and quantum evolution equations only differ by 
terms of higher order in $\hbar$. Moreover, for potentials quadratic in 
$\hat q$, higher order terms cancel out and the quantum evolution 
equation coincides with the classical one. 
This leads to the remarkable result that there is no difference between 
the evolution of the distributions of probability for quantum and 
classical observables, when the system is described by a Hamiltonian 
quadratic in positions and momenta. The generalization to $N$ particles and 
eventually to field theory is straightforward and may be found in ref. 
\cite{MRV98b}.

The scheme just outlined is selfconsistent and 
doesn't  require at any step external structures such as the wave 
function or the density matrix. In this sense the  probability description 
of quantum mechanics in terms of tomograms, once completely formulated,  
has to furnish a perfectly 
equivalent scheme to conventional ones \cite{MMT96}. Thus, it appears 
quite natural to investigate the symmetry properties of the MDF of identical 
particles with respect to  permutations.
To this regard we discuss for simplicity the case of two particles, while a 
generalization will be exhibited in the next section. 

From eq. \eqn{wro} the MDF of two
identical particles is represented by: 
\be
 w(\xi_1,\mu_1,\nu_1;\xi_2,\mu_2,\nu_2)=k_2 \int  D_2(z;z')
{\times} e^{ -i[\ap_1+\ap_2] } \rho(z_1,z_2;z_1',z_2';t)
\ee
where we have defined
\begin{itemize}
\item[i) ] $\ap_i=\frac{z_i-z_i'}{|\nu_i|\hbar}
\left[\xi_i-\mu_i\frac{z_i+z_i'}{2}\right]$,
\item[ii) ] $D_n(z;z';x;\ldots;y)=(\Pi_{i=1}^n dz_i)(\Pi_{i=1}^n dz'_i)
(\Pi_{i=1}^n dx_i)\ldots(\Pi_{i=1}^n dy_i)$,
\item[iii) ]$k_n\equiv \Pi_{i=1}^n\frac{1}{2\pi\hbar|\nu_i|}$;
\end{itemize}
and  the time-dependence has been omitted.
We note that because
\be
\rh(z_1,z_2;z'_1,z'_2)=\rh_1(z_1,z'_1)\rh_2(z_2,z'_2)
\ee
the tomogram of a system of two particles factorizes as: 
\be
w(\xi_1,\mu_1,\nu_1;\xi_2,\mu_2,\nu_2)=
w_1(\xi_1,\mu_1,\nu_1)w_2(\xi_2,\mu_2,\nu_2)
\ee
with
\be
w_i(\xi_i,\mu_i,\nu_i)=\frac{1}{2\pi\hbar|\nu_i|}\int dz_idz'_i
e^{-i\alpha_i}\rh_i(z_i,z'_i).
\ee
Therefore it seems  natural to identify the MDF relative to
symmetrized and antisymmetrized states as
\be
w_{\pm}(\xi,\mu,\nu)_2  :=  k_2\int  D_2(z;z')
e^{ -i[\ap_1+\ap_2] } \rho_\pm(z|z')_2  \label{eq:mdfpm}
\ee
with
$(\xi,\mu,\nu)_n\equiv (\xi_1,\mu_1,\nu_1;\xi_2,\mu_2,\nu_2;\ldots
\xi_n,\mu_n,\nu_n)$ (in the same way we may define $w_1$ and $w_2$ as 
the tomograms associated to $\rho_1$ and $\rho_2$.

From \eqn{eq:kern1}, which represents $\rho_{\pm}$ in terms of the integral 
kernel, we can write \eqn{eq:mdfpm} in the form
\be
w_\pm(\xi,\mu,\nu)_2 =  k_2 \int  D_2(z,y;z',y')
K_\pm(y,y'|z,z')_2 e^{ -i[\ap_1+\ap_2] } \rh(y|y')_2,
\ee
then, using the inverse formula \eqn{row} for the case of two particles
\be
\rh(y|y')_2 = \frac{1}{(2\pi)^2}\int D_2(m,x)
 e^{ i[\bt_1+\bt_2] } w(x,m,y-y')_2 \label{eq:ro2}
\ee
with  $\bt_i=x_i-m_i\frac{y_i+y'_i}{2}$
we obtain
\beqa
& & w_\pm(\xi,\mu,\nu)_2
=\frac{k_2}{(2\pi)^2}\int D_2(y,y',z,z',x,m,n)
\delta_{n_1,y_1-y'_1}\delta_{n_2,y_2-y'_2}  \nn \\
& & e^{ -i[\ap_1+\ap_2] }e^{ i[\bt_1+\bt_2] }
K_\pm(y,y;z,z')_2w(x,m,n)_2 \nn \\
& &=\int D_2(x,m,n)\Biggl\{\frac{k_2}{(2\pi)^2}\int
D_2(y,y',z,z')\delta_{n_1,y_1-y'_1}\delta_{n_2,y_2-y'_2} \nn \\
& & e^{ -i[\ap_1+\ap_2] }e^{ i[\bt_1+\bt_2] } K_\pm(y,y';z,z')_2
\Biggr\}w(x,m,n)_2 .
\eeqa
Thus the tomographic probability distribution corresponding to a
symmetric or antisymmetric state of two particles may be finally
written as
\be
w_\pm(\xi,\mu,\nu)_2=
\int D_2(x,m,n)\widetilde{K}_\pm\left(\xi,\mu,\nu|x,m,n\right)_2
w(x,m,n)_2 \label{eq:wpm}
\ee
with the kernel $\widetilde{K}_\pm$ given by
\beqa
\widetilde{K}_\pm(\xi,\mu,\nu|x,m,n)_2 &=&
\frac{k_2}{(2\pi)^2}\int D_2(y,y';z,z')
\delta_{n_1,y_1-y'_1}\delta_{n_2,y_2-y'_2}  \nn \\
&\times& \exp [ -i(\ap_1+\ap_2) + i(\bt_1+\bt_2) ]  K_\pm(y,y'|z,z')_2.
\label{eq:ker2}
\eeqa
Also in this case it is straightforward but not particularly illuminating 
to derive the integral kernels $ \widetilde{K}_1, \widetilde{K}_2$, 
associated to $w_1$ and $w_2$, which complete the realization of the 
permutation group on the space of tomograms.  

\section{Generalization to $N$ particles}
The generalization to N particles is now straightforward: we define the
kernel $K_\pm(y,y';z,z')_n$ as the following:
\beqa
K_\pm(y,y';z,z')_n &:=&\frac{1}{(n!)^2}
\left( 1+\sum_P \epsilon_P P_{j_1\ldots j_n}\right)
\otimes \left( 1+\sum_{P'} \epsilon_{P'} P'_{j_1\ldots j_n}\right)\nn\\
&\times& \dd{1}{1}\ldots\dd{n}{n}\ddd{1}{1}\ddd{n}{n}
\eeqa
where $\epsilon_P$ and $\epsilon_{P'}$ are equal to 1 in the symmetric 
case, while representing, in the antysymmetric case,  the sign of the
permutations $P$ and $P'$ respectively.
Therefore we obtain 
\be
\widetilde{K}_\pm(\xi,\mu,\nu)_n=
\frac{k_n}{(2\pi)^n}\int D_n(y,y',z,z')\left[\Pi_{j=1}^n
\delta_{n_j,y_j-y'_j}  e^{ -i\ap_j }e^{ i\bt_j }\right]
 K_\pm(y,y';z,,z')_n \label{KtildeN}
\ee
so that we find 
\be
w_\pm(\xi,\mu,\nu)_n=
\int D_n(t,m,n)\widetilde{K}_\pm\left(\xi,\mu,\nu|t,m,n\right)_n
w(x,m,n)_n. \label{wN}
\ee
We may  introduce a sort of generalized Slater determinant \cite{Sl30}
to obtain ${K}_\pm$, and, by means of \eqn{KtildeN}, 
${\widetilde K}_\pm$. We define
\be
\Delta_\pm(y|x)_n=\frac{1}{n!}
\begin{array}{||c c c c||}
\dd{1}{1} & \dd{1}{2} & \ldots & \dd{1}{n} \\
\dd{2}{1} & \dd{2}{2} & \ldots & \dd{2}{n} \\
\vdots & \vdots &  & \vdots \\
\dd{n}{1} & \dd{n}{2} & \ldots & \dd{n}{n} \\
\end{array}
\ee
with the convention that for the symmetric case we take always
the plus sign in computing the determinant.
In this way we can write the kernel $K_\pm(y,y'|z,z')$ in a more
conventional manner as:
\be
K_\pm(y,y';z,z')_n:=\Delta_\pm(y|z)_n\Delta_\pm(y'|z')_n.
\ee

Summarizing,  we have succeeded in realizing the symmetric and 
antisymmetric tomographic probability distribution associated to a 
system of $N$ identical particles, eq. \eqn{wN}, in terms of the 
 integral kernel \eqn{KtildeN}. This result fits into the selfconsistent 
scheme outlined at the beginning of this section in the sense of yielding a 
realization of the permutation group on the space of tomograms which 
doesn't require additional structures to be defined. In the forthcoming 
section we will see in concrete how it works on an example and we will 
check the invertibility retriwing the well known expressions for the 
symmetric and antysimmetric density matrices associated to a system of 
oscillators.

\section{Application}
Let us consider as an example the case of two independent harmonic 
oscillators. For this system the time-evolution equation \eqn{evo} has been 
solved in ref. \cite{MRV98a} and the solutions have the following 
expression: 
\be
w_{nm}(x,\mu,\nu)_2=\frac{e^{-y_1^2-y_2^2}}{\pi|r_1r_2|n!m!2^{n+m}}
H^2_n(y_1)H^2_m(y_2) \label{wnm}
\ee
where $r_j=e^{it}(\mu_j+i \nu_j)$, $y_j=x_j/|r_j|,~j=1,2$
and $H_n(y)$ and $H_m(y)$ are the Hermite polynomials of order
$n$ and $m$ respectively.

For the two-dimensional case the kernel may be obtained in a closed form by 
integrating  Eq.  (\ref{eq:ker2}) over the $z,z'$ and the $y,y'$ variables.
After some lengthy calculation we are able to write
\beqa
\widetilde{K}_\pm(\xi,\mu,\nu)_2 &=&
f(\xi_1,\mu_1,\nu_1;\xi_2,\mu_2,\nu_2)+
f(\xi_2,\mu_2,\nu_2;\xi_1,\mu_1,\nu_1) \nonumber \\
&\pm& \bigl[g(\xi_1,\mu_1,\nu_1;\xi_2,\mu_2,\nu_2)+
g(\xi_2,\mu_2,\nu_2;\xi_1,\mu_1,\nu_1) \label{Ktilde2}
\bigr]
\eeqa
with
\begin{itemize}
\item
$f(\xi_2,\mu_2,\nu_2;\xi_1,\mu_1,\nu_1)=\frac{k_2}{4}
\delta\left( m_1-\frac{\mu_1 n_1}{\nu_1}\right)
\delta\left( m_2-\frac{\mu_2 n_2}{\nu_2}\right)\exp\left\{i\left(
x_1+x_2-\frac{n_1}{\nu_1}\xi_1-\frac{n_2}{\nu_2}\xi_2\right)\right\}$;
\item
$g(\xi_2,\mu_2,\nu_2;\xi_1,\mu_1,\nu_1)=\frac{k_2}{2\pi}
\frac{\nu_1\nu_2}{|\mu_2\nu_1-\mu_1\nu_2|}
\exp{\left\{\frac{1}{2}v^T {\ensuremath{\cdot}} A{\ensuremath{\cdot}} v+B
{\ensuremath{\cdot}}v+i(x_1+x_2)\right\}}$;
\item
$v=\{m_1,n_1,m_2,n_2\}$;
\item
$A=\left(
\begin{array}{cccc}
i\frac{\nu_1\nu_2}{(\mu_1\nu_2\ - \mu_2\nu_1)} &
i\frac{\mu_2\nu_1 + \mu_1\nu_2}{2(\mu_2\nu_1 - \mu_1\nu_2)} & 0 & 0 \\
i\frac{\mu_2\nu_1 + \mu_1\nu_2}{2(\mu_2\nu_1 - \mu_1\nu_2)} &
i\frac{\mu_1\mu_2}{(\mu_1\nu_2\ - \mu_2\nu_1)}  & 0 & 0 \\
0 & 0 &  i\frac{\nu_1\nu_2}{(\mu_1\nu_2\ - \mu_2\nu_1)}  &
  i\frac{\mu_2\nu_1 + \mu_1\nu_2}{2(\mu_2\nu_1 - \mu_1\nu_2)}  \\
0 & 0 &  i\frac{\mu_2\nu_1 + \mu_1\nu_2}{2(\mu_2\nu_1 - \mu_1\nu_2)} &
  i\frac{\mu_1\mu_2}{(\mu_1\nu_2\ - \mu_2\nu_1)}
\end{array}
\right)$;
\item
$B=\left\{i\frac{\nu_1\xi_2-\nu_2\xi_1}{\mu_1\nu_2-\mu_2\nu_1},
i\frac{\mu_2\xi_1-\mu_1\xi_2}{\mu_1\nu_2-\mu_2\nu_1},
i\frac{\nu_1\xi_2-\nu_2\xi_1}{\mu_1\nu_2-\mu_2\nu_1},
i\frac{\mu_2\xi_1-\mu_1\xi_2}{\mu_1\nu_2-\mu_2\nu_1}
\right\}$
\end{itemize}
where $v^T$ denotes the transposed vector and $\ensuremath{\cdot}$ is the
usual matrix product, rows by columns. 
This is all what we need to obtain the symmetrized and
antisymmetrized MDF corresponding to all the states of the two oscillators,
although Eq. \eqn{eq:wpm} is in general not easy to integrate.

For the first non trivial case, i.e. the first excited
state, $w^{01}_\pm(\xi,\mu,\nu)_2$, we obtain, substituting Eq. \eqn{Ktilde2} 
and Eq. \eqn{wnm} for $w^{01}$ into Eq. \eqn{eq:wpm}   
\beqa
w^{01}_\pm(\xi,\mu,\nu)_2
&=& \int D_2(x,m,n)\widetilde{K}_\pm\left(x,\mu,\nu|x,m,n\right)_2
    \frac{e^{-y_1^2-y_2^2}}{\pi|r_1r_2|n!m!2^{n+m}}
    H^2_0(y_1)H^2_1(y_2) \nn \\
&=& \frac{ e^{ -\frac{\xi_1^2}{\mu_1^2+\nu_1^2}
               -\frac{\xi_2^2}{\mu_2^2+\nu_2^2}
             }
         }{ \pi\sqrt{ (\mu_1^2+\nu_1^2)(\mu_2^2+\nu_2^2) }
          }
\Biggl[
     \left( 2\frac{\xi_2}{ \sqrt{\mu_2^2+\nu_2^2} } \right)^2  +
     \left( 2\frac{\xi_1}{ \sqrt{\mu_1^2+\nu_1^2} } \right)^2
     \nn \\
&\pm&
     2\xi_1\xi_2\left(
   \frac{\mu_1\mu_2+\nu_1\nu_2}{(\mu_1^2+\nu_1^2)(\mu_2^2+\nu_2^2)}
          \right)
\Biggr].
\eeqa
As we can see, only the first two terms of the sum are probability 
distributions, whereas the last one is an interference term.

For our edification we can check, using the inverse formula
\eqn{eq:ro2},
that we obtain the correct expression for the symmetrized and 
antisymmetrized density matrices associated to a system of two harmonic 
oscillators, that is
\be
\rho^{01}_\pm(x,x')_2=\frac{1}{4}\exp{ \left\{ -\frac{1}{2}
(x_1^2+x_2^2+{x'_1}^2+{x'_2}^2) \right\} }\left[  4x_2 x'_2+4x_1x_1'\pm
(4x_1x_2'+4x_2x_1' )\right].
\ee
\section{Conclusions}
In summary, we have studied the permutational symmetry of tomograms of
quantum states for systems of identical particles of bosonic and
fermionic nature. We showed that the tomographic probability
distributions of identical particles in the probability representation
of quantum mechanics are associated to the realization of the action of
$G {\otimes} G$, where $G$ is the permutation group and ${\otimes}$
denotes the direct product. It is demonstrated that the tomographic
probability distribution, that is the MDF is a sum of terms, some of
them directly recognizable as tomograms, some others which are 
interference terms, that is not associated to a probability
distribution by themselves, but only in the given combination. The main
results of our analysis are summarized by Eqs. \eqn{KtildeN}, \eqn{wN},
where tomograms of identical particles are explicitely derived
in terms of an integral kernel. 
\section{Acknowledgments}
V. I. Man'ko thanks the University of Naples 
and INFN Sezione di Napoli
for kind hospitality and 
the Russian Foundation for Basic Research for partial support under 
project n. 01-02-17745.

\end{document}